\documentclass[aps,pre,twocolumn,amsmath,amssymb,superscriptaddress,floatfix,showpacs]{revtex4}
\usepackage{graphicx}
\usepackage{bm}
\input epsf.sty
\begin{document}
\title{Mobility and asymmetry effects in one-dimensional rock-paper-scissors games}
\author{Siddharth Venkat}
\author{Michel Pleimling}
\email{michel.pleimling@vt.edu}
\affiliation{Department of Physics, Virginia Polytechnic Institute and State University, Blacksburg, Virginia 24061-0435, USA}

\begin{abstract}
As the behavior of a system composed of cyclically competing species is strongly influenced by the presence of fluctuations,
it is of interest to study cyclic dominance in low dimensions where these effects are the most prominent.
We here discuss rock-paper-scissors games on a one-dimensional lattice where the interaction rates and the mobility can be
species dependent. Allowing only single site occupation, we realize mobility by exchanging individuals of different species.
When the interaction and swapping rates are symmetric, a strongly enhanced swapping rate yields
an increased mixing of the species, leading to a mean-field like coexistence even in one-dimensional systems.
This coexistence is transient when the rates are asymmetric, and eventually only one species will survive.
Interestingly, in our spatial games the dominating species can differ from the species that would dominate in the corresponding
nonspatial model. We identify different regimes in the parameter space and construct the corresponding dynamical phase
diagram.
\end{abstract}


\maketitle

\section{Introduction}
Evolutionary game theory \cite{Hof98,Now06} 
has led in the past to novel and interesting insights into the complex behavior that can emerge in
a multispecies ecological system (see for example \cite{Bus79,Sza07,Fre09} and references therein).
Cyclic dominance of competing "species" is not restricted to biological systems, but can also be found 
in economic and social systems. This competition can lead to a large variety of phenomena, ranging from the
emergence of regular spatio-temporal pattern to chaotic dynamics. Intrinsic fluctuations and omnipresent
nonlinearities yield a rich variety of scenarios, the two extreme cases being ever-lasting biodiversity or rapid
dominance of a single species and extinction of the others.

The three species rock-paper-scissors game 
\cite{Fra96a,Fra96b,Pro99,Sza01,Tse01,Ker02,Sza04,Kir04,Rei06,Rei07,Rei08,Cla08,Pel08,Rei08a,Ber09}
is one of the simplest cyclic dominance model where the influence
of fluctuations can be studied systematically. On the level of the mean-field rate equations this system
is characterized by a reactive fixed point, corresponding to species coexistence 
where every species yields one third of the total population. When adding fluctuations, one has
to distinguish between spatial and nonspatial systems. In a nonspatial system, the presence of
stochastic fluctuations yields as the final state one of the three absorbing states where only one species survives
whereas the other two have become extinct \cite{Rei06,Ber09}. Interestingly, the 'weakest' species, i.e. the less efficient
predator, survives in the case of asymmetric interactions \cite{Ber09,Fre01}. In a spatial system without diffusion,
species extinction takes place in one dimension \cite{Fra96a,Fra96b,Pro99}, whereas in two dimensions spatio-temporal 
pattern can emerge in the form of rotating spirals \cite{Pro99,Tse01,Rei07,Rei08,Pel08,Rei08a}. Recent studies of 
two-dimensional lattice models have revealed that the mobility of individuals,
realized in the form of an exchange of individuals, destabilizes the spirals 
for larger swapping rates, thus leading to species extinction in the long-time limit \cite{Rei08,Pel08,Rei08a}.

Real-world examples discussed in terms of rock-paper-scissors models range from coral reef invertebrates \cite{Bus79}
to lizard populations in California \cite{Sin96} and from competing bacterial strains \cite{Ker02,Kir04}
to self-organizing Min proteins \cite{Loo08}, to name but a few examples. It is worth noting that some of
the recent theoretical studies of low-dimensional rock-paper-scissors games have been motivated by the intriguing bacteria experiments.

It follows from our brief discussion of the recent work on systems with cyclic dominance
that stochastic fluctuations have important, and sometimes surprising, effects on this type of systems. 
Motivated by this observation, we wish to fully elucidate the role of fluctuations in low-dimensional
systems composed of cyclically dominating species.
We therefore propose in the following
to complement this line of research by studying numerically one-dimensional systems of mobile individuals, with species
dependent interaction and swapping rates. As we shall show, biodiversity is made possible in one dimension
through high mobility, which is different to the behavior in two dimensions where an increased mobility
leads to species extinction.
In addition, and in contrast to the nonspatial case, we discover for asymmetric interaction
and swapping rates that the law of the weakest is not strict in the spatial case and that it depends on
the system parameters whether the weakest survives or dies out.

Another motivation for our study of one-dimensional systems comes from the recent studies of bacterial populations
in nanofabricated landscapes (see for example \cite{Key06} where linear arrays of coupled microscale patches of habitat
are discussed). These investigations open the intriguing possibility of a future study of competing bacterial populations
in quasi one-dimensional systems.

Our paper is organized in the following way. After having introduced our model in Section II, we discuss in Section III the
case of symmetric, i.e. species independent, rates and show that a high mobility can lead
in one dimension to the emergence of species coexistence. In Section IV we allow for asymmetric, i.e. 
species dependent, interaction and swapping rates and show that this asymmetry leads to an interesting
dynamical phase diagram not observed in the nonspatial case. Finally, we conclude in Section V.

\section{One-dimensional rock-paper-scissors games}
In our spatial rock-paper-scissors games we consider three species living on a one-dimensional support that
are competing in a cyclic way. Calling $A$, $B$, and $C$ the three species, these interactions
can be cast in the following reaction scheme:
\begin{eqnarray}
A+B,~ B+A & \overset{k_{ab}}{\longrightarrow} & A+A \\
B+C,~ C+B & \overset{k_{bc}}{\longrightarrow} & B+B\\
C+A,~ A+C & \overset{k_{ca}}{\longrightarrow} & C+C
\end{eqnarray}
where we allow for species dependent interaction rates $k_{ab}$, $k_{bc}$, $k_{ca}$. In the following we
restrict ourselves to the case where every site is occupied by exactly one individual, such that interactions
only take place between nearest neighbors. In this setting mobility is realized by swapping the positions of
individuals on neighboring sites \cite{Rei08,Pel08}
\begin{eqnarray}
A+B & \overset{s_{ab}}{\rightleftarrows} & B+A \\
B+C & \overset{s_{bc}}{\rightleftarrows} & C+B\\
C+A & \overset{s_{ca}}{\rightleftarrows} & A+C
\end{eqnarray}
where the swapping rates $s_{ab}$, $s_{bc}$, $s_{ca}$ can again be species dependent.
Note that mobility can also be realized through simple diffusion steps if empty sites are allowed and/or
more than one individual can occupy a given lattice site \cite{Fra96b,Pel08,He09}. 

In our numerical simulations, every lattice site is initially occupied by any of the three species with the
same probability 1/3. The total number of individuals,
$N_A + N_B +N_C = N$, is a conserved quantity. Here $N_X$ is the number of individuals of species $X$ in the system and $N$ is the
number of lattice sites. We consider periodic boundary conditions and use sequential dynamics. Having selected a neighboring pair
of individuals belonging to species $A$ and $B$, we allow for the following events: $A$ dominates $B$ 
with probability $k_{ab}$, $A$ and $B$ exchange positions with probability $s_{ab}$, or no 
interaction takes place with probability $1 - k_{ab} - s_{ab}$. After having selected $N$ such pairs the time 
is increased by one unit.

First studies of one-dimensional systems with $M$ competing species have been 
published some time ago \cite{Fra96a,Fra96b,Pro99,Fra98}. If the individuals are immobile,
exact results can be obtained \cite{Fra96a,Fra96b}. For example,
it is found that the individuals organize in single-species domains whose average size $\langle \lambda
\rangle \sim t^\alpha$ grows
algebraically with time. The exponent $\alpha$ governing this domain growth depends on the number of competing 
species as well as on the chosen dynamics (parallel or sequential dynamics). For the case most relevant to
our study ($M=3$ and sequential updates) the exact result is $\alpha = 3/4$. Interestingly,
for parallel dynamics an asymptotic equivalence was established between the diffusion-reaction
approach and the lattice model with immobile individuals \cite{Fra96b}. Recent simulations of lattice models 
with sequential dynamics, where diffusion
is made possible through multiple occupancy of a site and/or the presence of empty sites \cite{He09},
also indicate that inclusion of simple diffusion does not lead to a qualitative different behavior.
However, as we show in the following, this is no longer the case when the mobility of individuals is realized
through exchanges.

\section{Symmetric interaction and swapping rates}
Let us start our discussion of mobility effects in one-dimensional rock-paper-scissors games
by first looking at the symmetric case where all rates are species independent. We therefore set in the
following $k= k_{ab} = k_{bc} = k_{ca}$ and $s= s_{ab} = s_{bc} = s_{ca}$, with $k + s = 1$.

\begin{figure}[t]
\centerline{\epsfxsize=3.80in\ \epsfbox{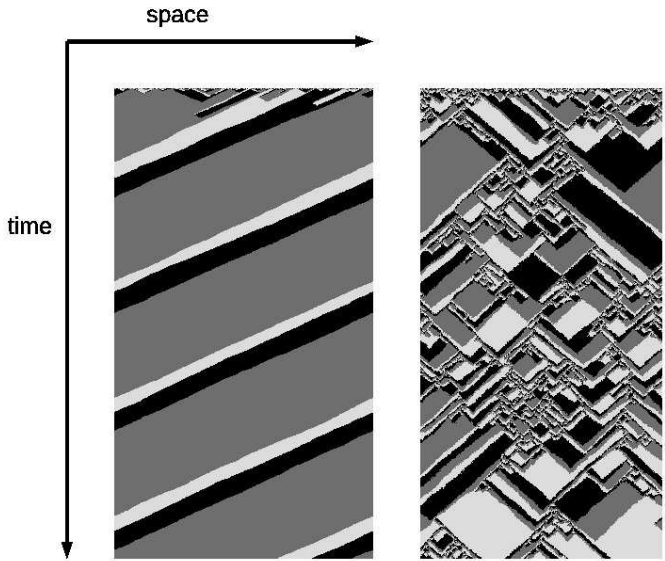}}
\caption{Space-time plots of one-dimensional rock-paper-scissors games with symmetric interaction rates $k$
and swapping rates $s$. White, gray, and black correspond to the three different species.
In the left panel, the individuals have a low mobility, with $s = 0.1$ and $k =0.9$,
yielding segregation of species, similar to what is observed for immobile individuals. In the right panel,
high mobility, with $s =0.9$ and $k = 0.1$, continuously mixes the different species, such giving raise to
coexistence even in one dimension. The system size is 2500 and the first 10000 time steps are shown.}
\label{fig1}
\end{figure}

\begin{figure}[t]
\centerline{\epsfxsize=3.20in\ \epsfbox{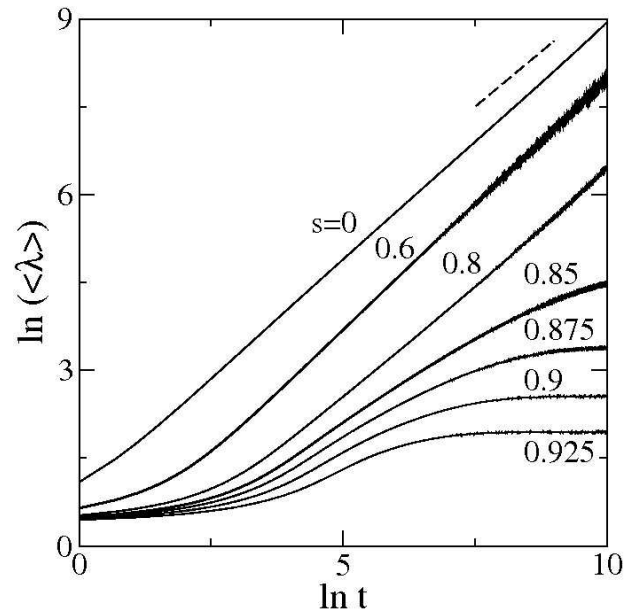}}
\caption{Average domain size as a function of time for different swapping rates $s$, with $k+s=1$.
The results shown here have
been obtained after averaging over 40000 independent runs for systems composed of 160000 individuals.
The dashed line indicates the exactly known result that for $s =0$ the time-dependent average domain size increases
algebraically with an exponent 3/4 for large times.}
\label{fig2}
\end{figure}

As already mentioned, earlier studies \cite{Fra96a,Fra96b,Pro99,Fra98} of immobile individuals
have highlighted the existence of species segregation through the formation of single-species domains 
that grow algebraically in time. If one allows for mobile individuals, the conclusions drawn in \cite{Fra96a,Fra96b}
remain unchanged for not too high swapping rates $s$. As an example we show in the left panel of Fig. \ref{fig1}
the space-time diagram for a system composed of 2500 individuals where the rates are $k=0.9$ and $s = 0.1$. Segregation sets in
immediately, yielding the formation of numerous small single-species domains. These domains then coarsen, and 
the system rapidly decomposes into a few large domains. At this stage, swapping has become
irrelevant and the behavior of the system is determined by the movement of the interfaces between different clusters.
The time evolution of the system is then similar to the time evolution of the system composed
of immobile individuals, yielding the same exponent $\alpha = 3/4$ for the average domain size, see Fig. \ref{fig2}. Eventually,
this process leads to the extinction of two of the three species as one species dominates the others in the long-time limit.

\begin{figure}
\centerline{\epsfxsize=2.60in\ \epsfbox{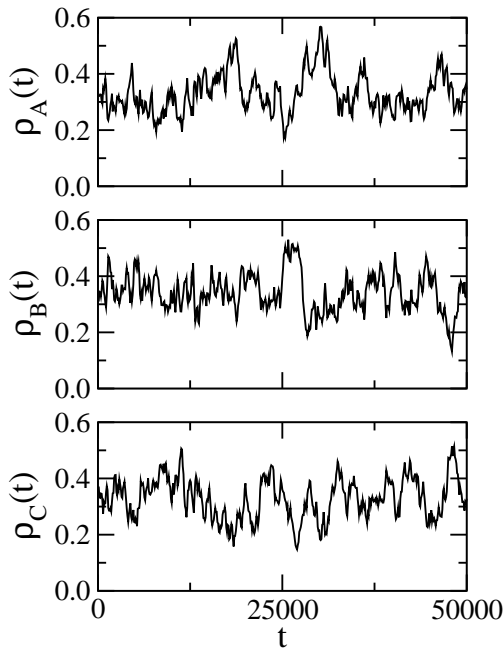}}
\caption{Time-dependent particle densities for all three species for a typical run with $k=0.1$ and $s=0.9$ and $N=10000$
lattice sites. In this regime, where $s > s_c$, an irregular oscillatory behavior is observed.}
\label{fig3}
\end{figure}

If one keeps increasing the mobility of the individuals, one eventually reaches a critical swapping rate $s_c = 0.84(1)$ above which 
the exchange mechanism very effectively mixes the different species. This yields completely different space-time plots, see the
right panel of Fig. \ref{fig1}, with a mosaic of ordered domains that emerge at some time and vanish again later. In addition to
these regular pattern one also observes more chaotic space-time regions where no large domains are formed. This intriguing
spatial-temporal behavior also shows up in the average domain size, see Fig. \ref{fig2}, as $\langle \lambda \rangle$ does no longer 
increase algebraically with time for $s > s_c$, but saturates at a finite, $s$ dependent, value. 
In this regime the particle
densities of all three species display an irregular oscillatory behavior, see Fig. \ref{fig3}. Obviously, the mixing 
induced by the high exchange rates
promotes the coexistence of species, and therefore biodiversity, in our one-dimensional system. 

This observation of coexistence in an ecological system of highly mobile individuals agrees with the
intuitive picture that a mixing of species should lead to a mean-field like behavior. This is different to
the more complex situation in two space dimensions where for medium values of the mobility an intermediate
regime exists in which rotating spirals are destabilized, yielding
rapidly a uniform state
where only one species survives. A coexistence regime similar to that observed by us in one space dimension eventually emerges 
in two dimensions for very high
mobilities \cite{Rei08,Pel08,Rei08a}.

Remarkably, a comparison of our results with the few existing results with simple diffusion (made possible through the 
presence of empty sites and/or sites with multiple occupancy) \cite{Fra96b,He09} indicates that the mechanism through
which mobility is realized matters. At this stage we can only speculate on the origin of these unexpected differences.
We note that in a field-theoretical description our exchange mechanism yields a nonlinear term, in contrast to simple
diffusion which enters linearly \cite{Tau09}. It is, however, beyond the scope of this article to explore this further
using field-theoretical techniques.

\section{Asymmetric interaction and swapping rates}
There are usually no compelling reasons that require all interaction rates to be the same for a system with competing species.
On the contrary, one should expect to have species dependent rates in any real system, as has also been
observed in a recent study of the self-organization
of Min proteins \cite{Loo08}. Theoretical studies usually neglect this aspect and exclusively focus on species independent
interaction rates. One exception of which we are aware of is the study of small asymmetries in a four-state
rock-paper-scissors game (where the forth state means that a site is unoccupied) in two dimensions \cite{Pel08}. 
For that model, it was shown through 
first-order perturbation theory that rotating spirals are robust against small asymmetries. 
Similar conclusions where drawn recently from a four-state rock-paper-scissors model that was constructed to
describe the symbiosis between an ant-plant and two protecting ant symbionts \cite{Szi09}.
In nonspatial rock-paper-scissors
games an asymmetry in the reaction rates was shown to lead to a 'law of the weakest', as the less efficient predator 
always survives \cite{Ber09}.

In principle, we have to explore for our three species model a six-dimensional parameter space. We will in the following 
not try to present a comprehensive study of all possible cases. Instead we fix the interaction and swapping rates between (1) $B$ and $C$
and (2) $C$ and $A$ particles and systematically change the rates for the processes involving $A$ and $B$ particles.
Concretely, we set $k_{bc} = k_{ca} = 0.4$ and $s_{bc} = s_{ca} = 0.4$.
In fact, this two-dimensional slice through our six-dimensional parameter space already captures our main result, 
namely that the 'law of the weakest' is not a strict one in spatial games.

\begin{figure}
\centerline{\epsfxsize=3.40in\ \epsfbox{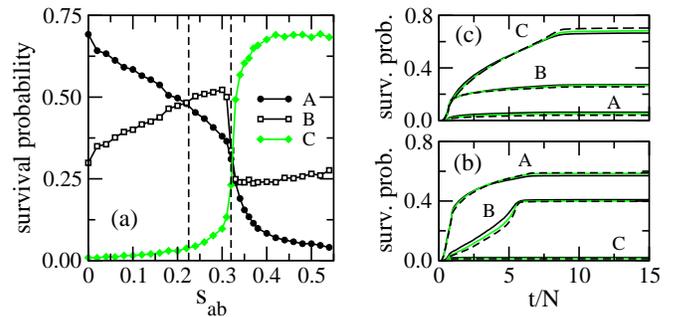}}
\caption{(Color online) (a) Survival probability of the different species as a function of $s_{ab}$ for $k_{ab} = 0.45$. 
The total population is 5000 individuals. The data result from averaging over 16000 different runs with 
different realizations of the noise. The dashed lines indicate the locations of the two transitions, see main
text. The errors are comparable to the size of the symbols.
(b) Survival probability as a function of time for $k_{ab} = 0.45$ and $s_{ab} = 0.10$. The different curves
correspond to different system sizes $N$, with $N=2500$ (full black lines), $N=5000$ (grey [green] lines), 
and $N=10000$ (dashed lines). (c) The same as (b), but now for $k_{ab} = 0.45$ and $s_{ab} = 0.50$.}
\label{fig4}
\end{figure}

We first remark that an asymmetry in the rates yields the dominance of a single species both in the spatial
and in the nonspatial game.
Whereas in nonspatial games the weakest predator, i.e. the species $X$ for which the interaction rate
$k_{xy}$ on another species $Y$ is smaller than the other interaction rates, always survives,
this is different in one space dimension. We discuss in Fig. \ref{fig4} the survival probabilities for the fixed
interaction rate $k_{ab} = 0.45 > k_{bc}$, $k_{ca}$. Writing the survival probability of species $A$ in a total population of $N$
individuals as $P_A(N)$, one readily verifies that the nonspatial game yields for this situation 
$P_B(N) \longrightarrow 1$ and $P_C(N), P_A(N) \longrightarrow 0$ in the macroscopic limit $N \longrightarrow \infty$,
i.e. the species $B$, which is dominated by $A$ and dominates $C$, survives.
In the one-dimensional case, however, the situation is much more complex, as the surviving species changes as
a function of the value of the exchange rate $s_{ab}$. As shown in  Fig. \ref{fig4}a for a system with 5000 individuals, 
we can identify three different regimes. Whereas for small exchange rates $s_{ab} < s_1$ with $s_1 = 0.225(5)$
the species $A$ has the highest 
survival probability, this changes for larger values as then either species $B$ (for $s_1 < s_{ab} < s_2$ with 
$s_2 = 0.32(1)$) or species $C$ 
(for $s_{ab} > s_2$) prevails. 
The transition at $s_{ab} = s_1$ is a smooth one, as none of the survival probabilities change dramatically when
crossing the transition point. This is different for the transition at $s_{ab} = s_2$ which is characterized by a very abrupt
change of the survival probabilities. The latter transition has therefore typical characteristics of a discontinuous
transition, whereas the behavior at the former transition is reminiscent of that at a continuous transition.
We carefully checked the robustness of our results against a change of the number
of individuals, see 
Fig. \ref{fig4}b,c.

\begin{figure}
\centerline{\epsfxsize=3.20in\ \epsfbox{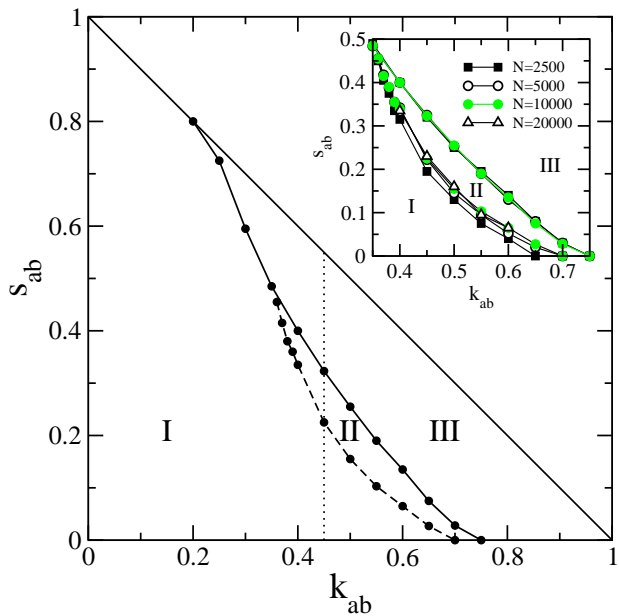}}
\caption{(Color online) Dynamical phase diagram for a two-dimensional slice through the six-dimensional parameter space, with
$k_{bc}= k_{ca} = 0.4$ and $s_{bc} = s_{ca} =0.4$. Three different phases characterized by the dominating species
(species $A$ in I, $B$ in II, and $C$ in III) are identified. Along the dashed resp. full line, one has the following
relation between the survival probabilities: $P_A = P_B > P_C$ resp. $P_A = P_B = P_C$. The vertical dotted line indicates
the line along which the data shown in Fig. \ref{fig4}a have been obtained. The inset shows the transition lines
for different system sizes.}
\label{fig5}
\end{figure}

Repeating this for other values of $k_{ab}$, we obtain the dynamical phase diagram shown in Fig. \ref{fig5}. 
Note that with our definition of the rates only the triangle defined by $k_{ab} + s_{ab} \leq 1$ is accessible. We
can distinguish three different phases which differ by the dominating species: in phase I species A dominates, in phase
II species B dominates, and in phase III species C dominates. These phases are separated by transition lines. The lower
transition line (dashed line) is characterized by the fact that $P_A = P_B > P_C$, whereas along the line separating phases II and III
we have $P_A = P_B = P_C$. As shown in the inset, the latter line does not show any size dependence. This is different for
the line separating the regimes I and II as here finite-size effects are observed for the smaller system sizes.
In fact, these finite-size dependences are compatible with a continuous transition along the line separating
regimes I and II and a discontinuous transition along the line separating regimes II and III.

Fig. \ref{fig5} seems to indicate the existence at $k_{ab} = 0.355(5)$ and $s_{ab}=0.47(1)$
of a triple point where the three phases meet. Based on the continuous and discontinuous character of the
different transition lines, this triple point should in fact be a tricritical point.
However, the existing data do not allow us to fully characterize the nature of this triple point,
due to subtle finite-size effects.
Systems much larger than those studied here might be needed in order to reliable determine the location
and the character of the triple point.

This complex dynamical phase diagram indicates that the law of the weakest observed in a nonspatial game is not strictly
valid any more for spatial systems. Thus, for $k_{ab} < 0.2$  we indeed observe that the most inefficient predator eventually
survives, as expected for the law of the weakest. However, the situation is much more complex for $k_{ab} >0.2$, as in addition
to the regime where the weakest survives one also encounters regimes where the weakest dies out.

\section{Conclusion}
Cyclic dominance in ecological or chemical systems is known to yield a very rich behavior and to lead to such intriguing 
features as rotating spirals. Very recently, a range of new insights have been gained through the study of three species
rock-paper-scissors games, and this in a spatial as well as in a nonspatial setting \cite{Rei06,Rei07,Rei08,Cla08,Pel08,Rei08a,Ber09}.

In this manuscript we complement this recent research activity by a numerical study of one-dimensional rock-paper-scissors games
with mobile individuals. We thereby realize the mobility through the exchange of neighboring individuals. In addition,
we allow for species dependent interaction and exchange rates, thereby bringing our model closer to real systems.

Interestingly, our study indicates that both the mobility and the rate asymmetry yield intriguing new features.

In presence of symmetric rates, a high mobility leads to a well-mixed system characterized by the prevailance of
coexistence in one dimension. The effect of the mobility is therefore different from the effect in two dimensions
where moderate mobility dissolves existing space-time pattern and ultimately leads to species extinction, whereas high
mobility leads to the emergence of coexistence \cite{Rei08,Pel08,Rei08a}, similar to what we observe. Another surprising
result is the observation that mobility realized through the exchange of individuals has a much stronger impact than
mobility realized through simple diffusion \cite{Fra96b,He09}.

For asymmetric, i.e. species dependent, rates we observe that the spatial support induces new phenomena not
observed in nonspatial games. Thus the law of the weakest, which states that the weakest predator survives and which is a
strict one in nonspatial games \cite{Ber09}, is replaced by a dynamic phase diagram with different regimes that are characterized
by the surviving species. A behavior in accordance with the law of the weakest can still be observed for certain 
ranges of the system parameters, but it is no longer a strict one.

We are not aware of any (quasi) one-dimensional experimental realization of the rock-paper-scissors games. A possible way
for a future realization of such a system could involve nanofabricated landscapes as those used in \cite{Key06} for a study
of bacterial populations. Our study points out the intriguing phenomena showing up in low dimensions and we hope that
these aspects will be addressed in future experiments.

In conclusion, our numerical study highlights the importance of species specific interaction and mobility rates in ecological systems.
Our results raise some new questions (the most intriguing being the observation that the effect of the mobility depends
on how it is realized and the possible existence of tricritical points in the dynamical phase diagram) 
and we plan to focus on these issues in the future.

\begin{acknowledgments}
We thank Uwe C. T\"{a}uber and Royce K. P. Zia for interesting and insightful discussions.
We are also grateful to Qian He, Mauro Mobilia, and Uwe C. T\"{a}uber for showing us their results
for systems with cyclic dominance and diffusion prior to publication.
This work was supported in part by the US National
Science Foundation through DMR-0904999. 
\end{acknowledgments}


\begin{thebibliography}{}
\bibitem{Hof98} J. Hofbauer and K. Sigmund, {\it Evolutionary Games and Population Dynamics} (Cambridge
University Press, Cambridge, UK, 1998).
\bibitem{Now06} M. A. Nowak, {\it Evolutionary Dynamics} (Harvard University, Cambridge, MA, 2006).
\bibitem{Bus79} L. W. Buss and J. B. C. Jackson, Am. Nat. {\bf 113}, 223 (1979).
\bibitem{Sza07} G. Szab\'{o} and G. F\'{a}th, Phys. Rep. {\bf 446}, 97 (2007).
\bibitem{Fre09} E. Frey, Lecture Notes for the 2009 Boulder School for Condensed Matter and Materials
Physics, http://boulder.research.yale.edu/Boulder-2009/ ReadingMaterial-2009/Frey/frey\_lecture\_notes\_games.pdf
\bibitem{Fra96a} L. Frachebourg, P. L. Krapivsky, and E. Ben-Naim, Phys. Rev. Lett. {\bf 77}, 2125 (1996).
\bibitem{Fra96b} L. Frachebourg, P. L. Krapivsky, and E. Ben-Naim, Phys. Rev. E {\bf 54}, 6186 (1996).
\bibitem{Pro99} A. Provata, G. Nicolis, and F. Baras, J. Chem. Phys. {\bf 110}, 8361 (1999).
\bibitem{Sza01} G. Szab\'{o} and T. Cz\'{a}r\'{a}n, Phys. Rev. E {\bf 63}, 061904 (2001).
\bibitem{Tse01} G. A. Tsekouras and A. Provata, Phys. Rev. E {\bf 65}, 016204 (2001).
\bibitem{Ker02} B. Kerr, M. A. Riley, M. W. Feldman, and B. J. M. Bohannan, Nature (London) {\bf 418}, 171 (2002).
\bibitem{Sza04} G. Szab\'{o} and G. A. Sznaider, Phys. Rev. E {\bf 69}, 031911 (2004).
\bibitem{Kir04} B. C. Kirkup and M. A. Riley, Nature (London) {\bf 428}, 412 (2004).
\bibitem{Rei06} T. Reichenbach, M. Mobilia, and E. Frey, Phys. Rev. E {\bf 74}, 051906 (2006).
\bibitem{Rei07} T. Reichenbach, M. Mobilia, and E. Frey, Phys. Rev. Lett. {\bf 99}, 238105 (2007).
\bibitem{Rei08} T. Reichenbach, M. Mobilia, and E. Frey, Nature {\bf 448}, 1046 (2007).
\bibitem{Cla08} J. C. Claussen and A. Traulsen, Phys. Rev. Lett. {\bf 100}, 058104 (2008).
\bibitem{Pel08} M. Peltom\"{a}ki and M. Alava, Phys. Rev. E {\bf 78}, 031906 (2008).
\bibitem{Rei08a} T. Reichenbach and E. Frey, Phys. Rev. Lett. {\bf 101}, 058192 (2008).
\bibitem{Sin96} B. Sinervo and C. M. Lively, Nature (London) {\bf 380}, 240 (1996).
\bibitem{Loo08} M. Loose, E. Fischer-Friedrich, J. Ries, K. Kruse, and P. Schwille,
Science {\bf 320}, 789 (2008).
\bibitem{Key06} J. E. Keymer, P. Galajda, C. Muldoon, S. park, and R. H. Austin,
Proc. Natl. Acad. Sci.  U.S.A. {\bf 103}, 17290 (2006).
\bibitem{Ber09} M. Berr, T. Reichenbach, M. Schottenloher, and E. Frey,
Phys. Rev. Lett. {\bf 102}, 048102 (2009).
\bibitem{Fre01} M. Frean and E. R. Abraham, Proc. Roy. Soc. {\bf 268}, 1323 (2001).
\bibitem{He09} Q. He, M. Mobilia, and U. C. T\"{a}uber, unpublished.
\bibitem{Fra98} L. Frachebourg and P. L. Krapivsky, J. Phys A {\bf 31}, L287 (1998).
\bibitem{Tau09} U. C. T\"{a}uber, private communication.
\bibitem{Szi09} A. Szil\'{a}gyi, I. Scheuring, D. P. Edwards, J. Orivel, and D. W. Yu,
Ecol. Lett. {\bf 12}, 1306 (2009).

\end{thebibliography}
\end{document}